\documentclass[twocolumn,showpacs,amsmath,amssymb]{revtex4}

    \usepackage{graphicx}
    \usepackage{subfigure}
    \usepackage{color}
    \usepackage{dcolumn}
    \usepackage{bm}
    \usepackage{xspace}

    \newcommand{\mss}{momentum superposition state\xspace}
    \newcommand{\wf}{wave function\xspace}
    \newcommand{\com}{center-of-mass\xspace}
    \newcommand{\qmr}{quantum mirror\xspace}

    \newcommand{\IgnoreThis}[1]{}
    \newcommand{\eins}{\mbox{$1 \hspace{-1.0mm} {\bf l}$}}
    \newcommand{\bact}{back-action}
    \newcommand{\qet}{entrain}

\begin{document}

\title{Imprinting interference fringes in massive optomechanical systems}

\author{Ole Steuernagel}

\affiliation{School of Physics, Astronomy and Mathematics,
University of Hertfordshire, Hatfield, AL10 9AB, UK }
\email{O.Steuernagel@herts.ac.uk}

\date{\today}

\begin{abstract}
  An interferometric scheme for the creation of {\mss}s of mechanical
  oscillators, using a quantum mirror kicked by free photons is
  analyzed. The scheme features ultra-fast preparation with immediate
  detection and should allow for the observation of signatures of
  momentum superpositions in a massive macros\-copic system at
  non-zero temperatures.  It is robust against thermalized initial
  states, displacement and movement, mirror imperfections, and the
  measurements' back-actions.
\end{abstract}


\maketitle

Heisenberg's uncertainty principle enforces that quantum measurements'
{\bact}s leave traces in an observed
system~\cite{Heisenberg_ZfP27,Bjoerk_PRA99}. Although their random
nature can be useful ({\bact} protects quantum cryptography protocols
from eavesdropping and it can help to cool tiny
mirrors~\cite{Kippenberg_SCI08}), the traces are usually detrimental
and {\bact} avoidance has been researched
intensively~\cite{BraginskyKhalili.book}. Uncontrollable measurement
{\bact}s give rise to loss of
coherence~(decoherence~\cite{Zurek_RMP03}) which hampers us when
building quantum computers, running sensitive interfero\-meters for
gravitational wave detection, or synthesizing superposition states of
classical objects.

In the thought experiment introduced here, we show that measurement
{\bact}s~\cite{BraginskyKhalili.book} can be restricted and harnessed
yielding a fruitful and stabilizing influence. Several probe particles
interact with a quantum system and are subsequently detected; the
traces they leave in the system modifies the future behaviour of
following probe particles. These repeated interactions can prepare the
system in a desirable quantum state and the features of that quantum
state can show up in modified measurement statistics of future probe
particles. An initially unbiased setup can thus become skewed by
repeated quantum interrogation. The system and its probe particles
have become {{\qet}ed}. This entrainment allows us to create otherwise
difficult-to-realize quantum states.

The ideas sketched here are related to work on relative localization
by Rau, Dunningham, and Burnett~\cite{Rau_SCI03,Dunningham-JMO04} and
follow-up work~\cite{Cable_PRA05,Douglas_arXiv1109.0041D}.

\begin{figure}[t]
\includegraphics[width=0.47\textwidth,height=0.42\textwidth]{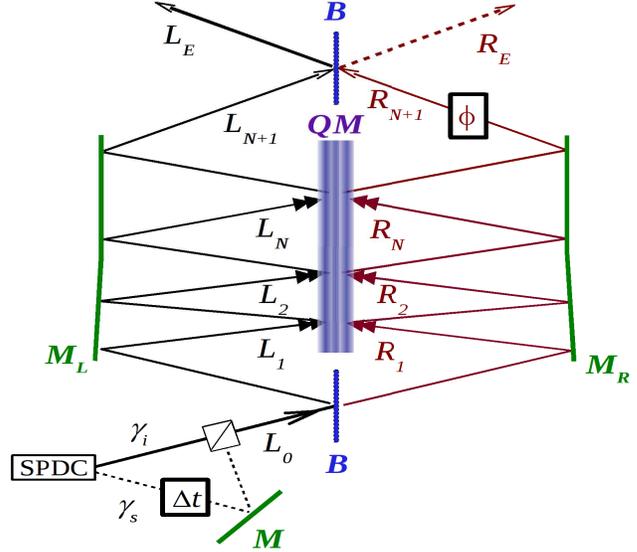}
\caption{(Color online) Setup for inter\-fero\-metric preparation and
  read-out of the state of a \qmr (QM). The initial photon~$\gamma_i$
  enters the inter\-fero\-meter through mode $L_0$, gets split into
  equal partial waves by a balanced beam splitter $B$ and traverses
  the inter\-fero\-meter via successive paths $L_1$, $L_2$, and so on
  (or alternatively via paths $R_1$, etc.). Every time it is reflected
  by the QM it imparts a momentum kick and thus prepares the mirror in
  a momentum superposition state (the corresponding modes are
  symbolized by folded double-arrows). A phase shifter $\phi$ allows
  us to scan the photons' interference patterns.  The final balanced
  beam mixer~$B$ removes `which-path' information; when the photon
  gets detected in mode $L_E$ or $R_E$, this measurement projects the
  mirror into a momentum-superposition state. With an ultra-short time
  delay, see Eq.~(\ref{eq_TimeDelay}), a second photon~$\gamma_s$
  follows~$\gamma_i$ via a polarizing beam splitter through the
  inter\-fero\-meter and interrogates the state of the QM.}
\label{Fig_setup_short}
\end{figure}

We consider a Michelson-Morley inter\-fero\-meter in which the
central, two-sided mirror is quantum delocalized in the $x$ direction,
perpendicular to its reflecting
surfaces~\cite{Kippenberg_SCI08,Marquardt_Physics.2.40,Aspelmeyer_NJP08},
see Fig.~\ref{Fig_setup_short}. The {\qmr}'s \wf is described by its
\com density matrix~$\rho(x,\xi)$ for which we want to assume that it
has a coherent extension of a few tens of nanometers (this might, for
example, be achieved through a ballistic expansion of a tightly
squeezed and cooled mirror~\cite{Corbitt_PRL07}
which is suddenly set free~\cite{Ole_arXiv1109.1818}).

In the first step of the entrainment procedure a single
photon~$\gamma_i$, such as those available from spontaneous parametric
down-conversion (SPDC) pair-creation processes~\cite{HongOuMandel87},
is sent through the inter\-fero\-meter, entering, say, through port~$L_0$.

In a classical inter\-fero\-meter, using a sharply localized perfect
mirror, the phase-shifter~$\phi$ can be set such that this photon
will exit through port~$L_E$ with certainty since destructive
interference renders port~$R_E$ dark. With a sufficiently widely
delocalized central mirror, however, this interference pattern gets
washed out and photons will exit through port~$R_E$ as well.

We want to concentrate on one photon at-a-time arrangements, the
next photon should interact with the mirror after the previous has
passed. The delay time between any two photons is therefore
constrained by
\begin{eqnarray}
\Delta t > \delta t + \frac{D (N-1)}{c} \; , \label{eq_TimeDelay}
\end{eqnarray}
here $\delta t \approx $ 100 fs is the photons' coherence
time~\cite{HongOuMandel87} and $D$ the distance they travel between
two mirror interactions. Note that for single bounce setups ($N=1$;
compare Fig.~\ref{Fig_setup_short}) the interaction time for $m$
photons is thus bound by $T_m\approx m \cdot \delta t$ and we can
generate and interrogate a momentum-superposition state repeatedly
on the pico\-second timescale.  This is in marked contrast to the
``standard approach'' of confining the light inside a
cavity~\cite{Marshall_Penrose_PhysRevLett.91.130401,Vitali_PRL07,Huang_PRA09,Jaehne_PRA09}.

For the formal analysis we need to determine the bosonic light-field
operators~$\hat L_E$ and~$\hat R_E$ at the exit ports in terms of
those at the entrance ports~$\hat L_0$ and~$\hat R_0$ (we will
leave~$\hat R_0 $ empty, see Fig.~\ref{Fig_setup_short})
\begin{eqnarray}
\begin{pmatrix}
 \hat L_E  \\
  \hat R_E
\end{pmatrix}
=  {\mathbf{B}}  {\mathbf{P}}_{N+1}
 {\mathbf{K}}_{N}   {\mathbf{P}}_{N}
\cdot
\ldots
\cdot
{\mathbf{K}}_{2}  {\mathbf{P}}_{2}
 {\mathbf{K}}_{1}  {\mathbf{P}}_{1}  {\mathbf{B}}
\begin{pmatrix}
  \hat L_0  \\
  \hat R_0
\end{pmatrix}
 \, .  \label{eq_field_mode_trafo}
\end{eqnarray}
The unitary 2$\times$2 matrices $ {\mathbf{B}}$,
$ {\mathbf{P}}$, and $ {\mathbf{K}}$ describe balanced mirrors,
photon propagators, and kick operators, respectively. Specifically,
$ {\mathbf{B}} =  {\mathbf{S}}(\frac{\pi}{4})$ is a special
case of a lossless splitter~$ {\mathbf{S}}$ with reflection
probability $\cos(\theta)^2$, namely
\begin{equation}
 {\mathbf{S}}(\theta) =
\begin{pmatrix}
  \cos(\theta) & i \sin(\theta)  \\
   i \sin(\theta) & \cos(\theta)
\end{pmatrix}
\, . \label{eq_beamsplitter_trafo}
\end{equation}
The photon propagators
\begin{equation}
 {\mathbf{P}}_j=
\begin{pmatrix}
 P_{L,j} & 0 \\ 0 & P_{R,j}
\end{pmatrix}
\end{equation}
account for the path length of mode ``$j$'' including the phase jump due
to the reflection by the perfect mirrors $M_L$ and $M_R$, respectively.

The kick operators enact the partial reflection and transmission of
photons by the quantum mirror in conjunction with the associated
momentum transfer to its \com density matrix~$\rho(x,\xi)$:
\begin{equation}
\hat{\mathbf{K}}_j(\theta) =
\begin{pmatrix}
  \cos(\theta) \, \hat K_{L_j}(\hat x) & i \sin(\theta) \otimes \eins \\
  i \sin(\theta) \otimes \eins  & \cos(\theta) \, \hat K_{R_j}(\hat x)
\end{pmatrix}
\; . \label{eq_kick_operator}
\end{equation}
With an angle of incidence~$\epsilon$ the effective photon momentum
transfer is $p_\gamma = 2 \hbar k \cos(\epsilon)$, where $k=2 \pi /
\lambda$ is their wave number and the kick operators in
Eq.~(\ref{eq_kick_operator}) have the form
\begin{eqnarray}
\hat{K}_{L_j} (\hat x) & = & \exp( \hat L_j^\dagger \hat L_j
\otimes \frac{i p_\gamma \hat x}{\hbar}) \; , \label{eq_kick_factors_L}
\\
\mbox{and } \quad \hat{K}_{R_j} (\hat x) & = & \exp(- \hat
R_j^\dagger \hat R_j \otimes \frac{i p_\gamma \hat x}{\hbar}) \; .
\label{eq_kick_factors_R}
\end{eqnarray}
The initial density matrix of the system (quantum mirror plus light
field) is
\begin{eqnarray}
    \varrho(x,\xi;l_0,r_0) = \rho(x,\xi) \frac{(\hat L_0^\dagger)^{l_0}
    (\hat R_0^\dagger)^{r_0} |0\rangle \langle 0| \hat L_0^{l_0}
    \hat R_0^{r_0}}{{l_0! r_0!}} \; .
 \label{eq_varrho}
\end{eqnarray}
We will from now on assume that only single photons are present
at a time (i.e., $l_0=1$ and $r_0=0$). The determination of photon
numbers at an output port of the inter\-fero\-meter involves tracing out
the quantum mirror and projecting onto that port (here,~$L_E$)
\begin{eqnarray}
  \langle \hat n_{L_E} \rangle =
  \langle \mathrm{Tr}_\mathrm{QM}\{  {\hat L^\dagger}_E \hat L_E  
  \varrho \} \rangle =
  \langle \int dx   \hat L_E^\dagger \hat L_E   \varrho(x,x)
  \rangle.\quad {}^{}
 \label{eq_NumberExpectVal}
\end{eqnarray}
Tracing over the field yields an effective kick operator~$\cal K$
acting on the mirror's density matrix~$\rho$. For example, for the
setup of Fig.~\ref{Fig_setup_short} with a single bounce off the
mirror ($N=1$) and assuming a photon enters through path $L_0$ and is
found to exit through port $L_E$ we have (with normal incidence
$\epsilon=0$)
\begin{eqnarray} {\cal K}_{L_E}
  & = & \left[ \sin(\theta)\cos(\frac{\phi}{2})-i\,
    \cos(\theta)\sin(2k x-\frac{\phi}{2}) \right]
  \nonumber \\
  & \times & \left[ \sin(\theta)\cos(\frac{\phi}{2})+ i\,
    \cos(\theta)\sin(2k \xi-\frac{\phi}{2}) \right] \, . \quad {}^{}
 \label{eq_KickOperator_NonUnitReflectivity}
\end{eqnarray}
For simplicity we write~$K_{L_E}=K_L$, then, similarly, ${\cal K}_{R}={\cal
  K}_{L}(\phi \mapsto \phi -\pi)$.

According to~Eq.~(\ref{eq_TimeDelay}) the time of interaction
between all successive photons and the mirror are very short, all
reference to the time evolution of the mirror is therefore absent in
our expressions for~${\cal K}$.

Since the \qmr's density matrix $\rho$ changes in response to the
port in which the exiting photon is detected, we re\-pre\-sent the
history associated with varying experimental outcomes through a
multi-index, namely, we write down the ports $L$ or $R$ in which the
exiting photons are registered:
\begin{eqnarray}
  \rho_{LRLL}(x,\xi) & = & ({\cal K}_{L} \rho_{RLL})(x,\xi) \nonumber \\
  & = & ({\cal K}_{L} {\cal K}_{R} {\cal K}_{L} {\cal K}_{L}
  \rho_{0})(x,\xi) \; ,
 \label{eq_rho_multiindex}
\end{eqnarray}
for example, describes the mirror's density matrix when the fourth
photon is seen in the left port after the first two were detected
there as well, but the third exited to the right.

The initial mirror density matrix $\rho_0$ is normalized: $\int dx
\; \rho_0(x,x) = 1$, this is not true for density matrices
conditioned on measurements. Only all conditional density matrices
taken together are normalized since, for $x = \xi$, we have
\begin{eqnarray}
{\cal K}_{L} + {\cal K}_{R} = 1
\, ,
 \label{eq_KickOperator_SUM_is_unity}
\end{eqnarray}
in other words, the integrated conditional density matrices carry
the relative weights for the occurrence of certain experimental
outcomes: $p_{H}=\int dx \; \rho_{H}(x,x)$. Here $H$ is the history
label which denotes the occurrence of a specific run, such as
$H=RLL$, in example~(\ref{eq_rho_multiindex}). We are thus led to
define the momentary spatial mirror probability density
\begin{eqnarray}
\alpha_{L,H}(x)  = \frac{\rho_{LH}(x,x) }{p_{H}} \; ,
 \label{eq_QMR_Prob_density}
\end{eqnarray}
which, when integrated over, yields the probability~$I$ to observe a
photon exiting through port~$L$ given a particular history~$H$
\begin{eqnarray}
I_{L,H}  = \int dx \; \alpha_{L,H}(x) \; .
 \label{eq_QMR_Probabiblity}
\end{eqnarray}
\begin{figure}[t]
\begin{center}
  \begin{minipage}[b]{0.32\linewidth}
    \includegraphics[width=0.99\linewidth,height=0.99\linewidth]{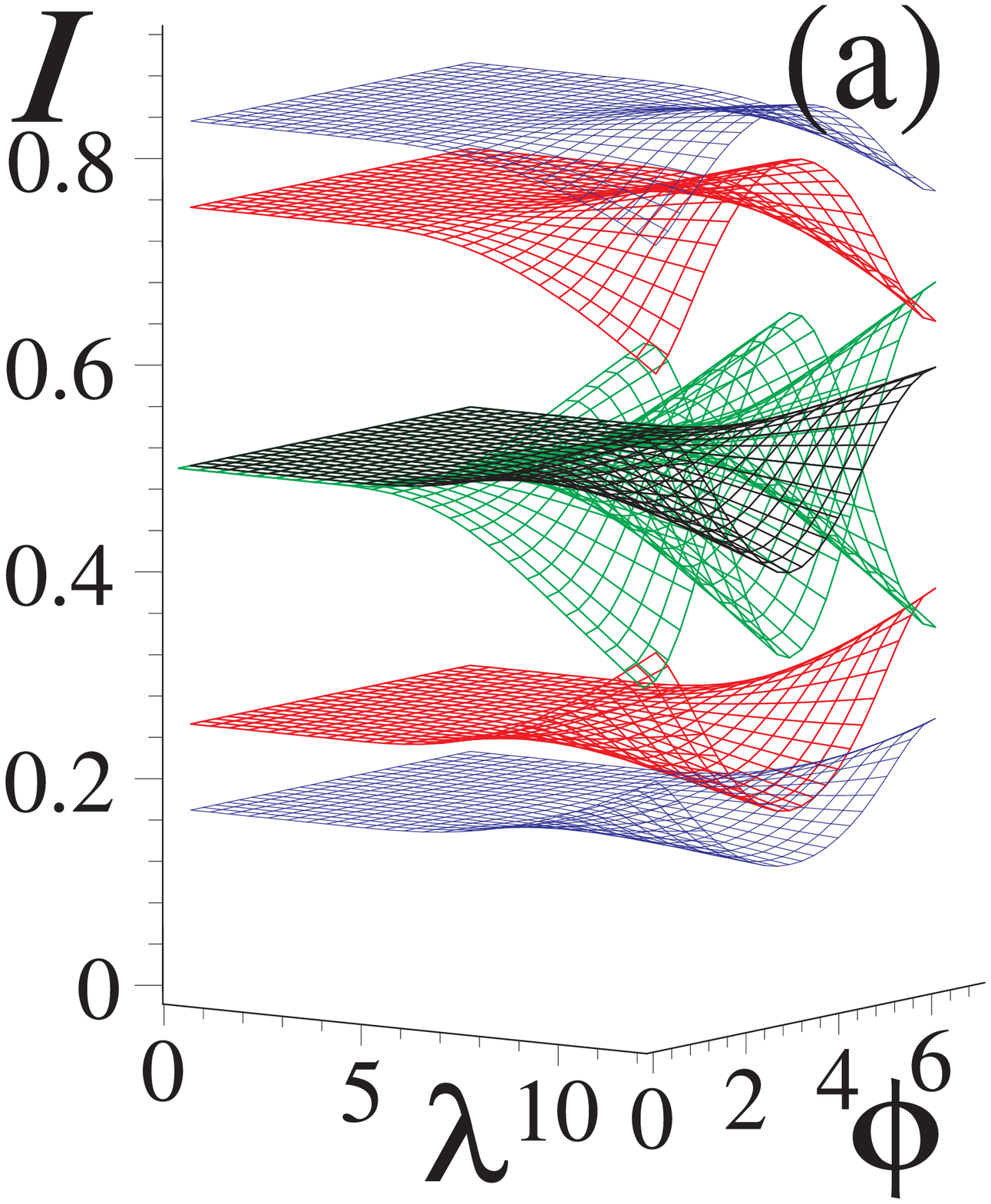}
 \end{minipage}
  \begin{minipage}[b]{0.32\linewidth}
    \includegraphics[width=0.99\linewidth,height=0.99\linewidth]{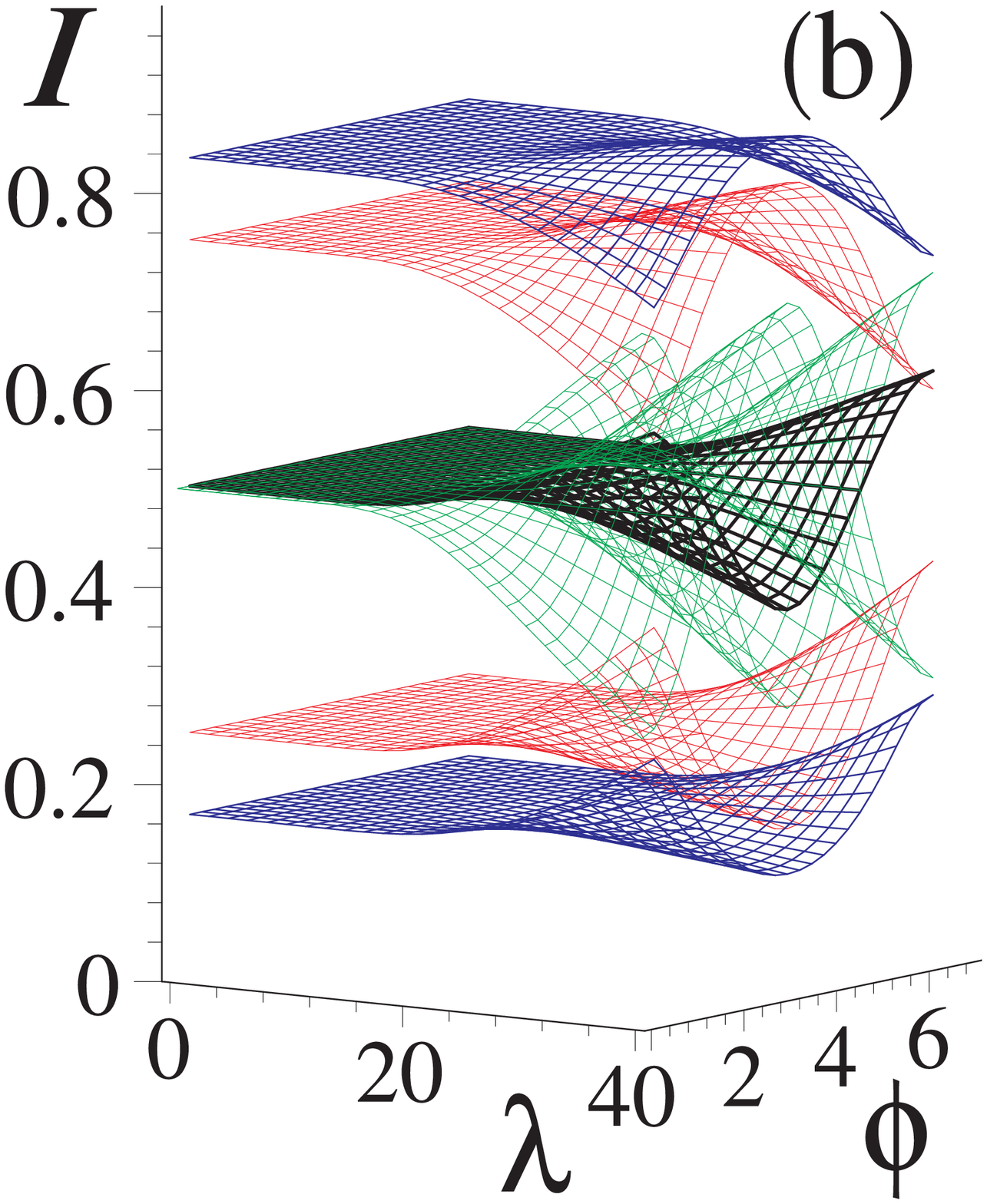}
 \end{minipage}
  \begin{minipage}[b]{0.32\linewidth}
    \includegraphics[width=0.99\linewidth,height=0.99\linewidth]{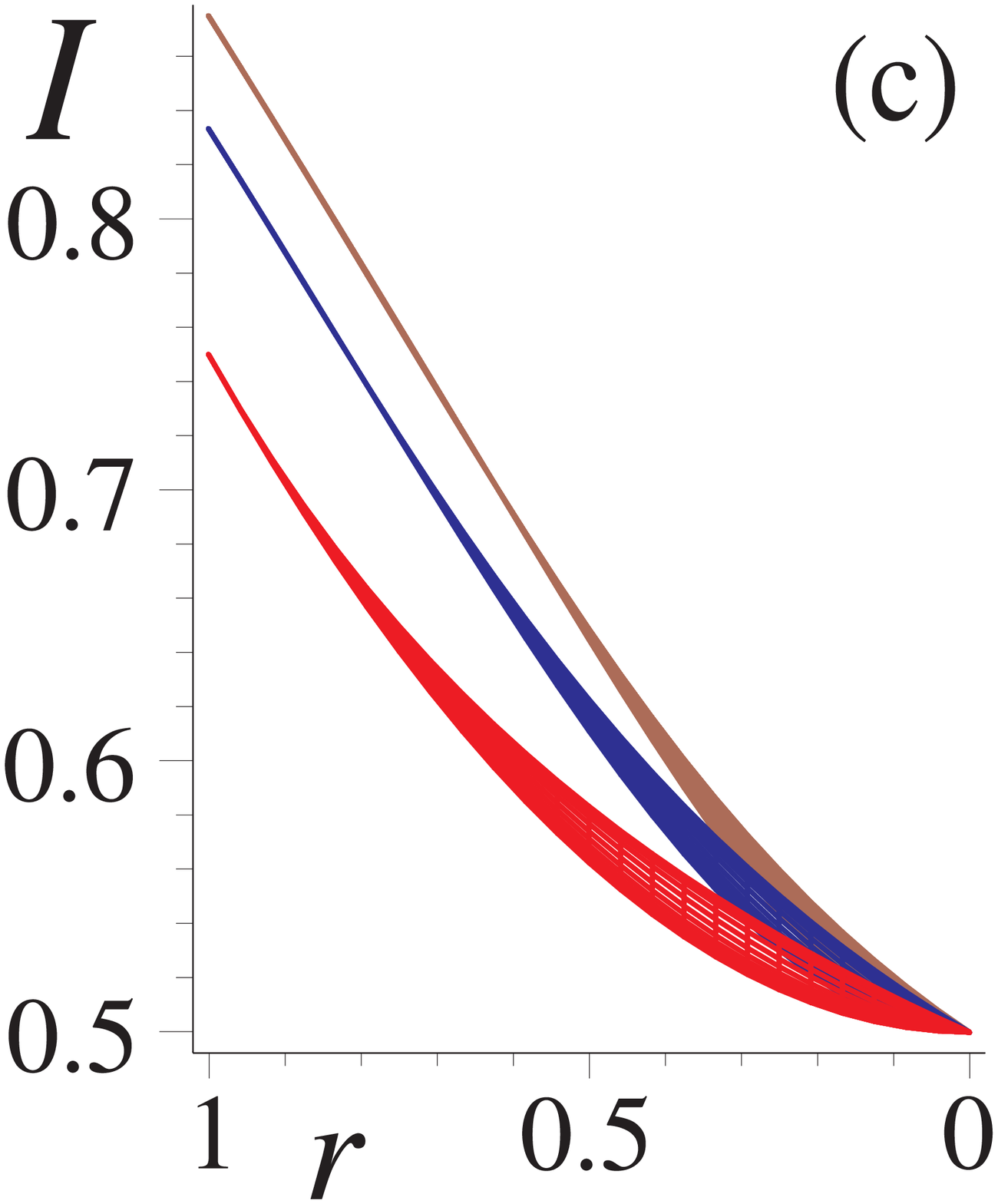}
 \end{minipage}
\end{center}
\caption{
  (Color online) Intensity distribution $I(\lambda,\phi)$ for
  single-bounce setup ($N=1$) of a Gaussian \qmr with perfect
  reflectivity $r=1$ [initial density matrix for a fully coherent
  state $\rho_0(x,x)=\exp(-x^2/\sigma^2)/ (\sigma\sqrt{\pi})$ with
  spread $\sigma=1$]. ({\bf a}) For small values of
  wavelength~$\lambda$ interference is washed out whereas for values
  of $\lambda/ \sigma > 6$ it shows: $I_{L}$ and $I_{R}$ (green sheets
  centered around 0.5). Detection of second $I_{L,L}$ (red sheet
  centered on 0.75) and third photon $I_{L,LL}$ (blue sheet centered
  on $5/6\approx 0.83$) shows strong {photon entrainment}. For mixed
  histories the weights are strongly reduced $I_{R,L}$ (red, at 1/4),
  $I_{R,LL}$ (dark blue, at 1/6) and $I_{R,LR}$ (thick black mesh
  centered on 0.5). ({\bf b}) Same plot as ({\bf a}) for triple-bounce
  case $N=3$. The effective resolution of the probe particles rises to
  $\Lambda \approx \lambda/(3\cdot 6)$: above $\lambda/ \sigma \approx
  18$ the quantum washout of the interference pattern
  diminishes. ({\bf c}) Same plot as ({\bf a}) for $I_{L,L}$,
  $I_{L,LL}$ and $I_{L,LLL}$, for ($\lambda \ll \sigma$), as a
  function of decreasing mirror reflectivity~$r$: the entrainment
  persists for imperfect mirrors. The curves' widths indicate small
  variations with change of the phase angle~$\phi$ ($\phi$-axis not
  shown).}
\label{fig_Densities_intensities_photons}
\end{figure}

Since the kick operators ${\cal K}$ depend on wavelength $\lambda$ and
phase setting~$\phi$, the intensity $I(\lambda,\phi)$ does as well,
compare Fig.~\ref{fig_Densities_intensities_photons}. Obviously
$I_{L,H}+I_{R,H}=1$, and for single-photon at-a-time scenarios $I$
equals the photon intensity $I=\langle n \rangle$ of
Eq.~(\ref{eq_NumberExpectVal}).

\begin{figure}[t]
\begin{center}
  \begin{minipage}[b]{0.32\linewidth}
    \includegraphics[width=0.99\linewidth,height=1.09\linewidth]{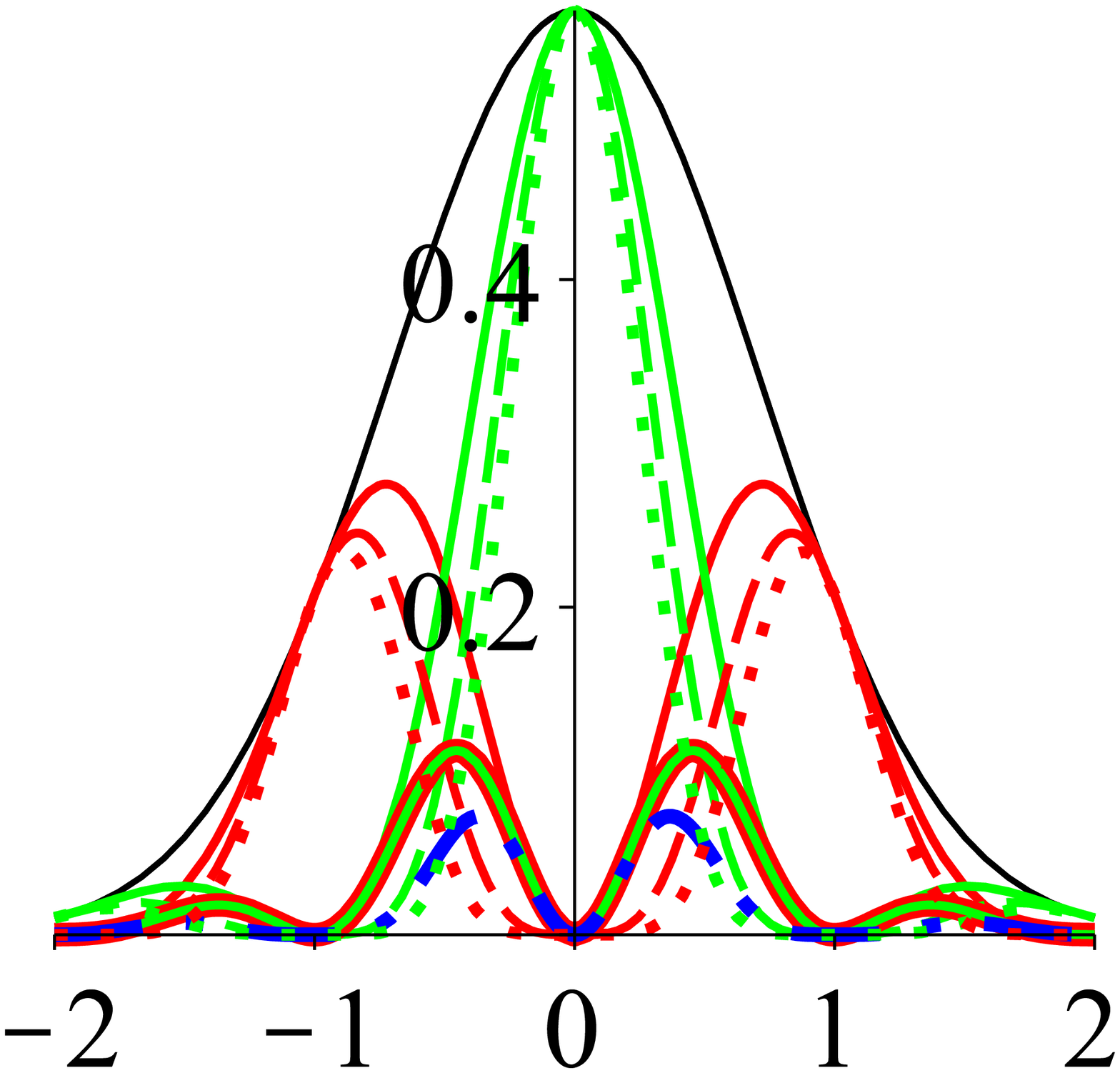}
 \end{minipage}
  \begin{minipage}[b]{0.32\linewidth}
    \includegraphics[width=0.99\linewidth,height=1.09\linewidth]{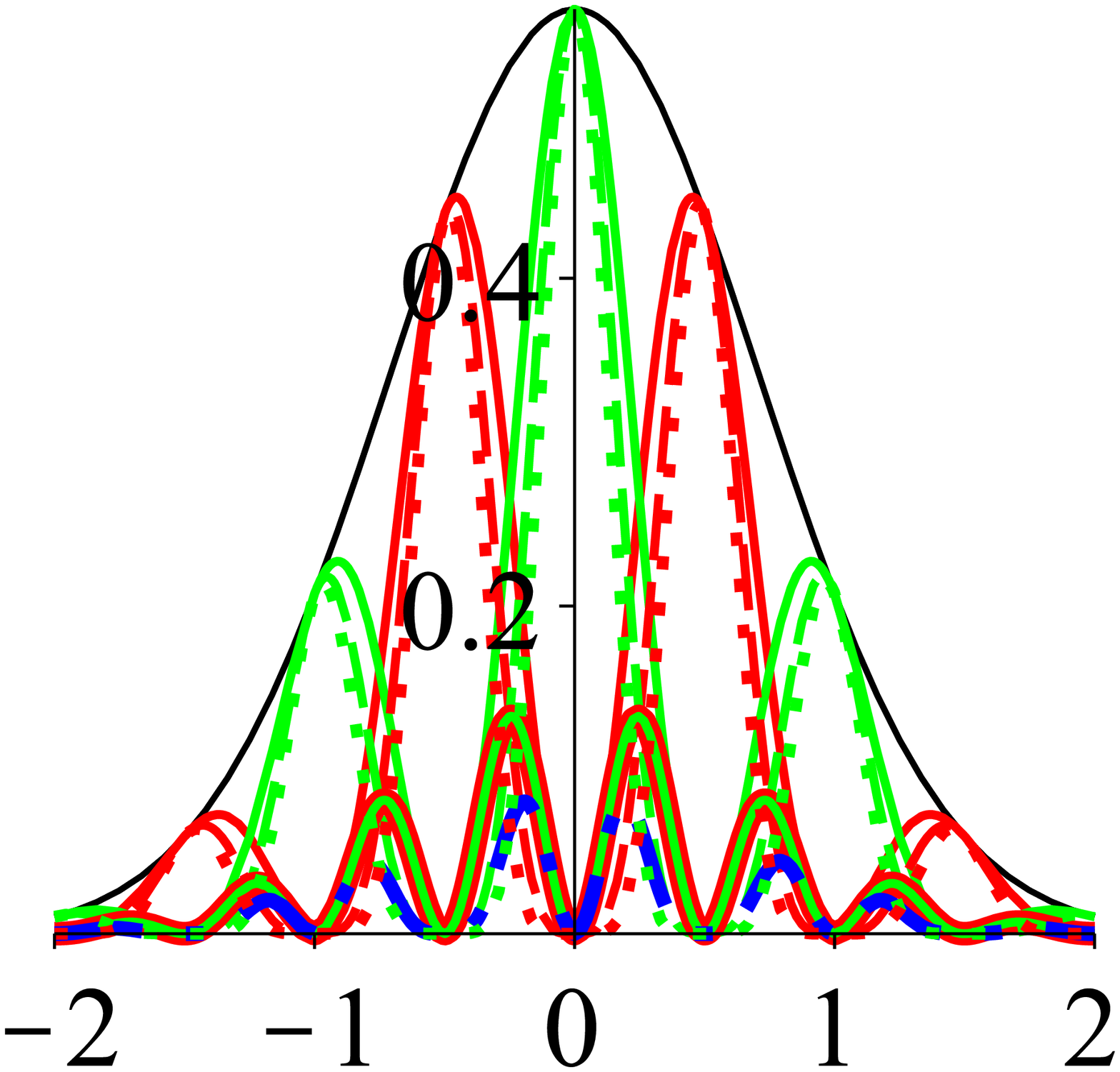}
 \end{minipage}
  \begin{minipage}[b]{0.32\linewidth}
    \includegraphics[width=0.99\linewidth,height=1.09\linewidth]{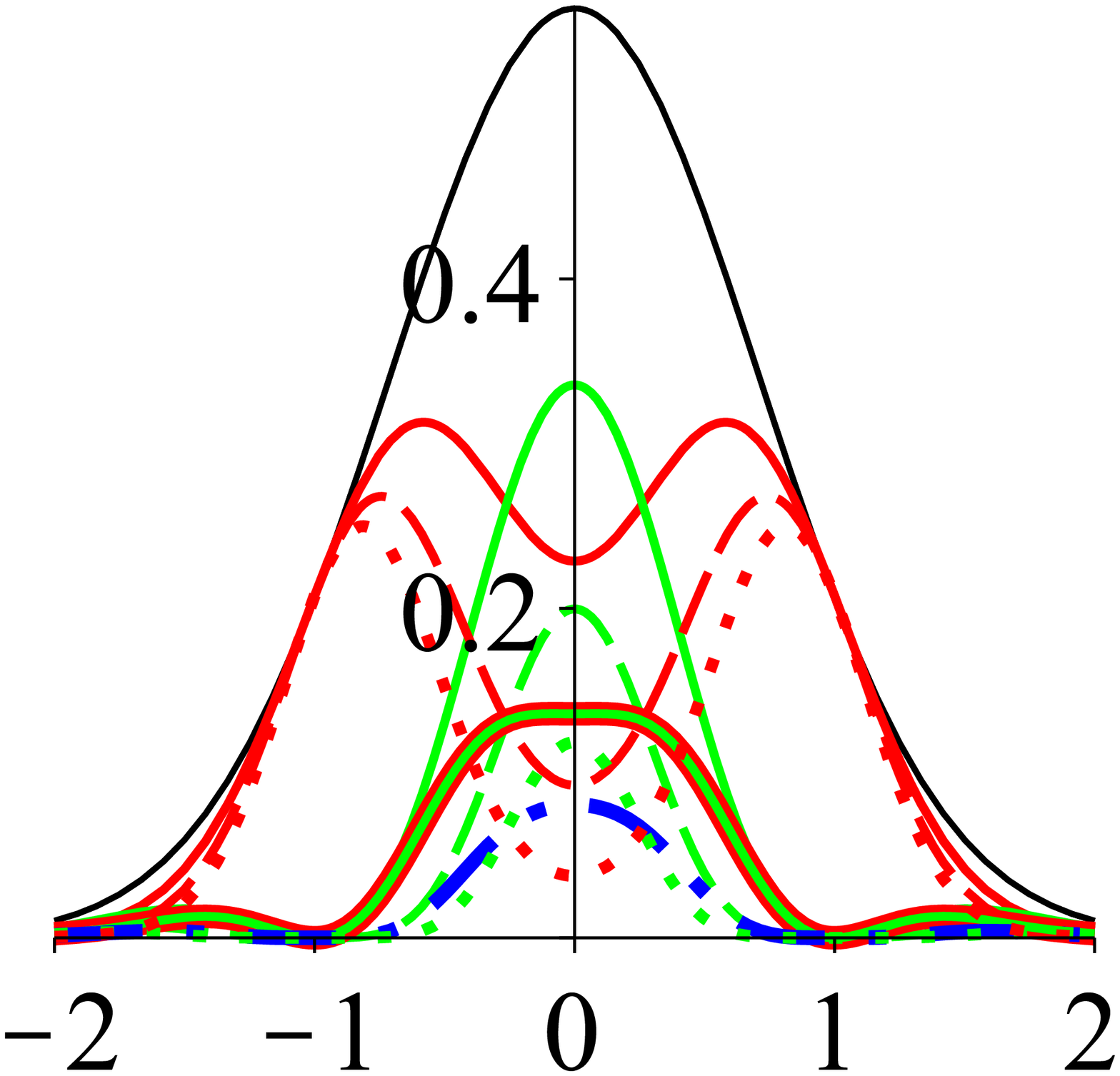}
 \end{minipage}
\end{center}
\caption{(Color online) Probability densities $\rho(x,x)$ of \qmr in
  initially Gaussian state with $\sigma=1$, $\lambda=1$ and symmetric
  setup, $\phi=0$: thin solid black envelope. ({\bf a}) Single-bounce
  setup ($N=1$), after the first photon has been detected: $\rho_{R}$
  and $\rho_{L}$ (solid single humped centered green line and solid
  double-humped red line); similarly after detection of second
  $\rho_{RR}$, $\rho_{LL}$ and third photon $\rho_{RRR}$, $\rho_{LLL}$
  (green and red dashed and dotted lines). For mixed measurement
  histories the weights are strongly reduced $\rho_{RL}=\rho_{LR}$
  (green-red superposed double-humped lines) and $\rho_{RLR}$ (blue
  dash-dotted line) this clearly demonstrates entrainment. ({\bf b})
  Double-bounce setup ($N=2$), compared to ({\bf a}) the imprint
  wavelength has halved. ({\bf c}) Same as ({\bf a}) for an imperfect
  mirror with reflectivity $r = 60\%$.}
\label{fig_Densities_rho_reflex100}
\end{figure}

The effective spatial wavelength~$\Lambda$ for imprint and
interrogation can be determined from
eq.~(\ref{eq_KickOperator_NonUnitReflectivity}) and is
\begin{eqnarray}
  \Lambda  = \left. \frac{\lambda}{4\cdot f \cdot N}
  \right|_{f_\mathrm{Gauss} \approx 1.5} \approx \frac{\lambda}{6 \cdot N}
  \; ,
 \label{eq_Imprint_Lambda}
\end{eqnarray}
where the form factor $f=1$ for a top hat and roughly 1.5 for a
Gaussian wave packet, this is best seen in
Fig.~\ref{fig_Densities_intensities_photons}~({\bf b}). This shrinkage
of the effective \emph{imprint} and \emph{interrogation wavelength}
$\Lambda$ is noteworthy, compare plots in
Figs.~\ref{fig_Densities_intensities_photons}
and~\ref{fig_Densities_rho_reflex100}.

The above kick factors are special cases of the general back-action a
photon imparts onto its scatterer. Typically its back-action destroys
coherence~\cite{Ole_PRA95}, but here the inter\-fero\-meter
geometrically restricts the photons to two (incoming and two
reflected) modes only. We therefore end up with the desirable
kick factors~$\cal K$ that represent controlled, quantum-superposed
momentum kicks. This allows us to create {\mss}s from initially
stationary \qmr states and allows for their detection and
reinforcement through {\qet}ment.

For a sufficiently wide mirror \wf we end up with sine- or
cosine-shaped imprint patterns for ${\cal K}_L$ or ${\cal K}_R$
respectively. Hence, $\rho_L$ and $\rho_R$ become approximately
orthogonal {\wf}s, a second photon~$\gamma_s$ picks up this trace and
tends to follow the first photon. This happens with roughly a 75\% :
25\% bias, see Fig.~\ref{fig_Densities_intensities_photons}, the
system has thus become {\qet}ed. The second photon's detection
moreover imprints the same kick factor onto the mirror's \com \wf thus
reinforcing this trend. The third and fourth photons follow their
predecessors with an increasing bias of roughly 83\% and 87\%,
respectively, see Fig.~\ref{fig_Densities_intensities_photons}~({\bf
  c}). Each time, the mirror gets kicked in an identical fashion this
procedure reinforces the interference fringes.

It should probably be emphasized that the sine- or cosine-shaped
imprint patterns are to be interpreted as our increase in knowledge
about the localization of the mirror according to classical wave
optics. Without further \emph{a priori} information about the nature of the
initial state of the mirror the method presented here does not allow
us to infer that an interference imprint has been created or
detected. The method works for any kind of mixed state and is
therefore fairly insensitive to temperature- and other effects, such
as non-zero average \com velocities and displacements~$\Delta x$ of
the average \com position of the \qmr, as long as~$\Delta x \ll c
\cdot \delta t$.

The rapidity of this method and the fact that it only probes the
mirror at chosen points in time reduces its contribution to
decoherence. For mirrors initially in sufficiently widely spread-out
pure states the back-action imprints and detects interference
imprints. An interrogation-photon's arrival time can be delayed to
allow for the investigation of the \qmr's time evolution and its
decoherence.

All features discussed above prevail for imperfect mirrors even when
their reflectivity drops to 60\% or less, see
Figs.~\ref{fig_Densities_intensities_photons}~({\bf c})
and~\ref{fig_Densities_rho_reflex100}~({\bf c}).

To conclude: An analysis of free photons interacting with a
quantum-delocalized mirror inside an inter\-fero\-meter shows that
their recoil can create and investigate {\mss}s of massive objects
non-destructively, within a pico\-second. The analysis makes use of
the entrainment of following photons by their predecessors. Such
entrainment may well turn out to be a useful new response mode of
quantum systems in various settings.


\end{document}